\documentclass[twocolumn,11pt]{aastex62}
\usepackage{ulem}
\usepackage{amsmath}
\usepackage{graphicx}
\usepackage{CJK}

\long\def\symbolfootnote[#1]#2{\begingroup%
\def\thefootnote{\fnsymbol{footnote}}\footnote[#1]{#2}\endgroup}

\newcommand{\beq}{\begin{equation}}
\newcommand{\eeq}{\end{equation}}
\newcommand{\bea}{\begin{eqnarray}}
\newcommand{\eea}{\end{eqnarray}}

\begin{document}

\title{\large \bf Polarization of kilonova emission from a black hole-neutron star merger}

\begin{CJK*}{UTF8}{gbsn}
\author{Yan Li (李彦) }\thanks{liyan287@mail2.sysu.edu.cn}
\affiliation{School of Physics and Astronomy, Sun Yat-Sen University, Zhuhai, 519082, P. R. China}
\author{Rong-Feng Shen (申荣锋)}\thanks{shenrf3@mail.sysu.edu.cn}
\affiliation{School of Physics and Astronomy, Sun Yat-Sen University, Zhuhai, 519082, P. R. China}

\begin{abstract}

A multi-messenger, black hole (BH) - neutron star (NS) merger event still remains to be detected. The tidal (dynamical) ejecta from such an event, thought to produce a kinonova, is concentrated in the equatorial plane and occupies only part of the whole azimuthal angle. In addition, recent simulations suggest that the outflow or wind from the post-merger remnant disk, presumably anisotropic, can be a major ejecta component responsible for a kilonova. For any ejecta whose photosphere shape deviates from the spherical symmetry, the electron scattering at the photosphere causes a net polarization in the kilonova light. Recent observational and theoretical polarization studies have been focused to the NS-NS merger kilonova AT2017gfo. We extend those work to the case of a BH-NS merger kilonova. We show that the degree of polarization at the first $\sim 1$ hr can be up to $\sim$ 3\% if a small amount ($10^{-4} M_{\odot}$) of free neutrons have survived in the fastest component of the dynamical ejecta, whose beta-decay causes a precursor in the kilonova light. The polarization degree can be $\sim$ 0.6\% if free neutrons survived in the fastest component of the disk wind. Future polarization detection of a kilonova will constrain the morphology and composition of the dominant ejecta component, therefore help to identify the nature of the merger.

\end{abstract} 

\section{Introduction}

It is well known that electron scattering of an unpolarized beam of incident radiation causes a partially linear polarization of the scattered radiation, whose polarization degree depends on the scattering angle.  For the radiation scattered into the plane perpendicular to the incident radiation, the polarization degree is the highest \citep{rybicki}. For a semi-infinite, plane-parallel, pure electron scattering atmosphere with a constant net flux, the polarization of its emergent light varies from a maximum of 11.7\% for an edge-on view to zero for a face-on view \citep{chandra}.


This leads to an important astrophysical implication. For a scattering dominated atmosphere and suppose it is angularly unresolved, there is a net polarization only when the visible portion of the atmosphere shows a deviation from the axisymmetry about the line of sight. Therefore, polarization detection becomes a tool to constrain the morphology of the emitting surface. A particular application is to the study of the aspherical nature of the supernova (SN) explosions.

\cite{hoflich91} calculated the polarization for pure scattering atmospheres of oblate or prolate spheroidal shapes using a Monte Carlo method, and compared with results observed from SN 1987A. Similar method was taken by \cite{kasen03} for type Ia SN 2001el. \cite{shapiro82} analytically calculated the polarization of those shapes, but restricting the line of sight to within the equator. Spectropolarimetry has now been carried out for supernovae of every major type and of various luminosity classes and peculiarities. They all show polarized light (the degree of polarization ranges from 0.2\% to 10\%) and hence asphericity in some significant way \citep{wang08}. 

A similar situation is a pure scattering, optically thin, non-spherical circumstellar envelope in which a star is centrally embedded, which was discussed analytically by \cite{brown77} and numerically by \cite{daniel80}. The polarization degree $\Pi_s$ for the scattered light from the envelope can be as high as $\sim 10\%$, but the net polarization of the total light (observed flux mainly comes directly from the central source) is  $\Pi_r \approx \Pi_s \tau_{\rm es} \sin^2 i$, where $i$ is the inclination angle of line of sight, and $\tau_{\rm es}$ is the scattering optical depth of the envelope which represents the fraction of the light to be scattered.  

True absorption shall always be present in the atmosphere. In an explosive scenario, as \cite{karp77} have discussed, the velocity gradient within the photosphere significantly enhances the bound-bound opacity as a result of the Doppler broadening of lines. The net effect of this Doppler broadening of many individual lines is the production of a quasi-continuous bound-bound opacity, which can be in magnitude comparable to that of electron scattering.

This is indeed the situation in the neutron-rich dynamical ejecta in NS-NS or NS-BH mergers which contains heavy element as r-process products, except that the bound-bound opacity there is more than two orders of magnitude higher than that of electron scattering (\cite{kasen13,tanaka13}; also see Figure 4 of \cite{metzger17}). The radioactive-decay heating of the ejecta by those unstable isotopes produces a so-called kilonova emission \citep{li98}. Therefore, for an average kilonova photon, its last interaction with matter before escaping the photosphere is more than likely a bound-bound transition, rather than an electron scattering. Consequently, the polarization from kilonovae might still be low, even though the ejecta shapes are very asymmetric \citep{kyutoku13, kyutoku14}.

As the kilonova counterpart to the NS-NS merger event GW170817, the optical to near-infrared observations of AT2017gfo represent the first opportunity to detect and scrutinize a sample of freshly synthesized r-process elements. \cite{covino17} made the polarization measurement of AT2017gfo at 1.46 days after GW 170817.  They report a null linear optical polarization degree of P = (0.50 $\pm$ 0.07)\%. However, as pointed out in \cite{covino17}, if the wind from the post-merger disk contributes a substantial mass for the kilonova ejecta \citep{just15,fernandez16}, the smeared and blended bound-bound opacity can be largely reduced because r-process in the wind ejecta is suppressed by neutrino emission \citep{kasen15}. This lanthanide-free wind component is predicted to give an earlier, `blue' kilonova component \citep{metzger17}. Because the wind terminal velocity depends on the local escape velocity from the disk surface, one expects that the photosphere of the wind component elongates toward the polar direction, as was illustrated in Figure \ref{fig:comp}, rendering a good condition for net polarization. Therefore, the hope for polarization detection resides with earlier observations.

Recently, \cite{bulla18} applied a Monte Carlo simulation to the NS-NS merger scenario and obtained a polarization degree similar to what was measured for AT2017gfo. \cite{matsumoto18} analytically estimated the evolution of the polarization for this situation. As pointed out in \cite{matsumoto18}, the existence of free neutrons in the outermost layer of the NS-NS merger ejecta is of great importance to detect early time  polarization. In this paper, we extend this line of work to the case of a BH-NS merger kilonova, which is expected to show more significant polarization than that of a NS-NS merger. Adopting the analytic model of \cite{shapiro82}, which is different from the Monte Carlo approach or result used by \cite{matsumoto18} and \cite{bulla18}, we can give detailed constraints not only on the mass of free neutrons but also the morphology and components of the ejecta.   

This paper is structured as follows. In section 2 we describe the ejecta components of a BH-NS merger in more detail. Our polarization calculation is described in section 3. The results are presented and analyzed in section 4. We summarize the results and discuss the implications in section 5.

\section{Ejecta components in a BH-NS merger}

\begin{figure}[h]
\begin{center}
\includegraphics[width=8cm, angle=0]{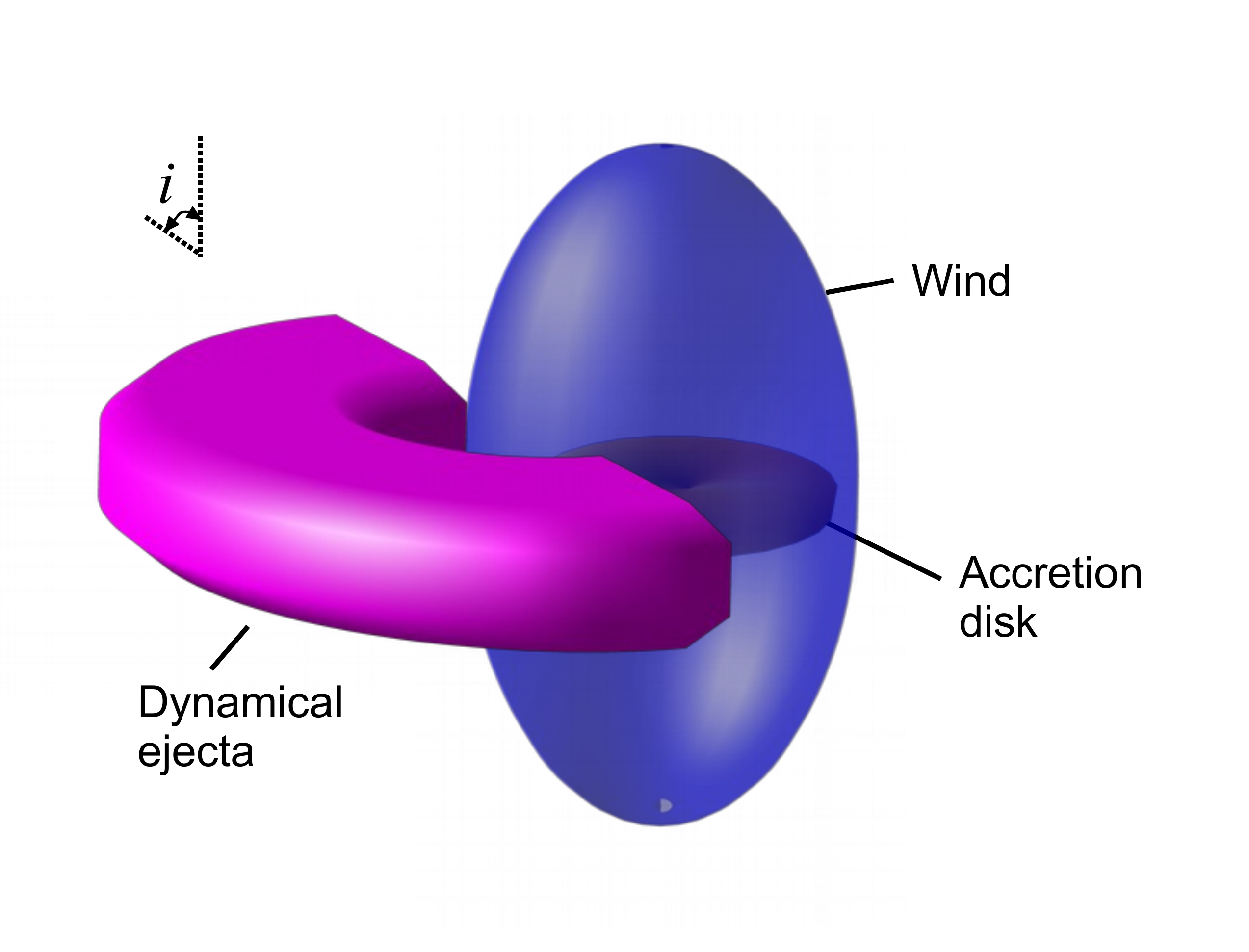}
\caption{The ejecta components in a BH-NS merger. The tidal tail ejecta of the disrupted NS is concentrated in the orbital plane, and within an azimuthal angle of $\sim \pi$ \citep{foucart14,kyutoku15,kyutoku18}. The magnetic/viscous/neutrino-driven disk wind component has low opacity and is elongated toward the polar direction (e.g., \cite{fernandez18}). In addition, a small amount of free neutrons (not shown here) may survive in the fastest, top layer of either the tidal ejecta or the disk wind \citep{metzger15}.}		\label{fig:comp}
\end{center}
\end{figure}

\cite{bhatta18} discussed various ejecta components (i.e., dynamical ejecta, disk wind, jet, cocoon) in BH-NS mergers and their detection prospects. They conclude that the kilonova is the most easily detected electromagnetic signature of all. Below, we consider two major ejecta components responsible for the kilonova -- the dynamical ejecta and the disk wind -- and a minor one, the free-neutron layer in the outermost layer of either one of the two major components. These components are illustrated in Figure \ref{fig:comp}.  

\subsection{Dynamical ejecta}

In a BH-NS merger, when the neutron star is disrupted prior to the merger, there is substantial matter ejected on the dynamical time scales (typically milliseconds) by tidal forces of the BH. The dynamical ejecta emerges primarily in the equatorial plane and shows significant anisotropy, different from a NS-NS merger whose dynamical ejecta is nearly isotropic \citep{Kawaguchi15, just15}. \cite{foucart14}, \cite{kyutoku15,kyutoku18} performed numerical-relativity simulations of BH-NS mergers to study dynamical mass ejection, and found that the dynamical ejecta is concentrated in the equatorial plane with a vertically half opening angle of $10^{\circ}-20^{\circ}$ and sweeps out about a half of the plane. The ejecta mass ranges from 0.01 to 0.1 $M_{\odot}$, and the average velocity of the ejecta inferred from the kinetic energy is typically 0.2 -- 0.3 c. This ejecta is highly neutron rich, producing heavy r-process elements that include Lanthanides which result in a high opacity $\kappa\sim 10$ cm$^2$/g \citep{kasen17}. 

\subsection{Disk wind}

The debris of the disrupted NS, that is bound to but not immediately swallowed by the BH, will circularize into an accretion disk. The disk mass is typically $\sim$ 0.01 -- 0.3 $M_{\odot}$, depending on the total mass and mass ratio of the binary, the spins of the binary components, and the NS EOS \citep{oechslin06}. Neutrino- and viscously- driven outflows from this remnant disk, taking place on longer timescales of up to seconds, provide another important source of ejecta \citep{just15, kiuchi15, fernandez18}. 

A high-resolution numerical-relativity simulation for the merger of a BH and a magnetized NS \citep{kiuchi15} showed that the amount of the disk wind component in the highest-resolution run is as high as $\sim  0.06 M_{\odot}$ which is much larger than that of the dynamical component $\sim 0.01 M_{\odot}$. Another simulation of BH-NS merger by \cite{{fernandez18}} that includes general relativity MHD effects showed that 40\% of the initial disk mass is ejected, with an average velocity of 0.1c and a non-spherical distribution, i.e., a hourglass shape, with an approximate 3:1 ratio between vertical and horizontal dimensions. 

\cite{just15} found that the major part of the disk wind material ends up in forming $A <130$ r-process elements, and the disk outflow complements the dynamical ejecta by contributing the lighter r-process nuclei $(A <140)$ that are underproduced by the strong r-process taking place in the very neutron-rich material expelled during the early coalescence phase. As a result, the opacity of the disk wind would be lower than that of the dynamical ejecta.
 
\subsection{Free neutron layer}	     \label{sec:n-component}

During its early expansion, most of the ejecta remains dense enough to allow the capture of all neutrons into nuclei via the r-process. However, some of the fastest ejecta may expand so rapidly that most neutrons contained in that head layer are not captured and become free \citep{metzger15}. The NS-NS simulation by \cite{fernandez18} found $\sim 10^{-4} M_{\odot}$ of the disk wind contains free neutrons. \cite{radice18,radice18b} simulated the turbulence-heating driven mass ejection during the NS-NS mergers and found that the fastest component ($\gtrsim 0.6~ c$ and close to $10^{-4} M_{\odot}$) is potentially free-neutron dominated.

The radioactive decay of these free neutrons (into protons with the emission of electrons) -- if their existence is robust -- can be an additional energy source of the kilonova emission, besides those of the dynamical ejecta and the disk wind. \cite{metzger15} calculated this precursor emission due to the extra heating by an outermost, free-neutron layer of the dynamical ejecta. As an ejecta expands, its diffusion time $t_{\rm diff}$ decreases with time as $\propto t^{-1}$, until eventually radiation can escape on the expansion timescale, at which $t_{\rm diff} = t$ (\cite{arnett82}). Therefore, for the outermost layer mainly containing free nucleons and electrons with a mass of $M_n$ and an expansion velocity of $v$, the characteristic timescale at which its light curve peaks is 
\begin{equation} 
\begin{split} 	\label{eq:tpeak}
{t_{\rm peak}} &\equiv {\left( {\frac{{M_n \kappa }}{{4\pi vc}}} \right)^{\frac{1}{2}}} \\
&\approx 0.63{\left( {\frac{M_n}{{{{10}^{ - 4}}{M_ \odot }}}} \right)^{\frac{1}{2}}}{\left( {\frac{v}{{0.5c}}} \right)^{ - \frac{1}{2}}}{\left( {\frac{\kappa }{{0.4 c{m^2}/g}}} \right)^{\frac{1}{2}}} hr,
\end{split}
\end{equation}
where the opacity $\kappa$ takes the value of electron scattering opacity for ionized hydrogen $\kappa_{\rm es,H} = 0.4$ cm$^2$ g$^{-1}$. 

Compared with NS-NS mergers, BH-NS mergers received relatively less attention. In the literature one finds fewer numerical simulations about mass ejection in BH-NS mergers, and very few paid particular attention to the fastest tail of the ejecta's mass-velocity distribution. In simulations by \cite{foucart14} and \cite{kyutoku15,kyutoku18}, the maximum velocity of the dynamical ejecta is $\sim$ 0.5 c, and a more asymmetric BH-NS mass ratio generally results in higher asymptotic velocities for ejetca. No simulation has been done particularly about the disk mass ejection following a BH-NS merger. But as was pointed out in \cite{fernandez18}, a disk in a BH-NS merger shall have the same physics and follow the same qualitative evolution as that in a NS-NS merger, whereas a larger central BH mass in the former case is expected to cause quantitative differences.

As the supporting evidence regarding the fastest ejecta components in a BH-NS merger is still inconclusive, here in this paper we still allow the possibility of an outmost layer of free neutrons with a mass of $\sim 10^{-5} - 10^{-4}$ $M_{\odot}$, in either the dynamical ejecta or the disk wind. Future observations could prove or disapprove this assumption.

\section{Polarization calculation}

Extensive studies have been done in the context of non-spherical SN explosions to estimate the degree of polarization. In particular, \cite{shapiro82} analytically calculated the polarization of oblate or prolate spheroidal atmospheres as a function of their asphericity parameter. Alternative approach of Monte Carlo calculation of photon propagation was taken by \cite{hoflich91} and \cite{kasen03}. \cite{brown77} have shown that the polarization degree is proportional to the sine squared of the inclination angle.

For a kilonova produced in a BH-NS merger, in the case that free neutrons survive in the outermost layer of the ejecta (the dynamical ejecta or disk wind), the electron scattering opacity dominates, thus a larger degree of polarization would be possible for a proper viewing angle. On the other hand, once the photosphere recedes to the bulk of the ejecta which is mainly made of r-process nuclei, the bound-bound transitions would dominate the opacity and result in a sharp decrease of polarization. 

Borrowing the insight from \cite{brown77} and \cite{matsumoto18}, the net polarization from an ellipsoidal shape of photosphere can be generally estimated as 
\begin{equation}	\label{eq:pinet}
{\Pi_{\rm net} \approx \Pi_0 \tau_{\rm es} \sin^2 i}
\end{equation}
where $i$ is the inclination angle of the line of sight with respect to the symmetry axis of the ellipsoid, $\Pi_0$ is the maximum degree of polarization that a given shape of ellipsoid could ever obtain, and $\tau_{\rm es}$ is the scattering optical depth of the photosphere; usually $\tau_{\rm es} < 1$ due to the presence of absorption. $\Pi_{\rm net}$ reaches its maximum value $\Pi_0$ only when $\tau_{\rm es}=1$ (i.e., pure-scattering dominated photosphere) and $i=90^{\circ}$ (i.e., seen sideways). The following two subsections describe the calculations of $\Pi_0$ and $\tau_{\rm es}$, respectively. The kilonova light curves are also calculated.

\subsection{The maximum degree of polarization $\Pi_0$}   \label{sec:pi0}


To estimate the maximum degree of polarization for an atmosphere of asymmetrical shape, we choose to use the following analytical formulas in \cite{shapiro82} because they contain a dependence on the asphericity of the atmosphere and they are easy to computate, rather than to adopt the Monte Carlo approach in \cite{daniel80}, \cite{hoflich91} and \cite{kasen03}.

For an semi-infinite, plane-parallel, optically thick, pure-scattering dominated atmosphere (a slab) as was considered by \cite{chandra}, the degree of polarization of the scattered emission along some direction $\mu$ is given by  
\begin{equation}
{\Pi (\mu)} = {{Q} \over {{I}}} = {{I_l- I_r} \over {{I_l + I_r}}} ,
\end{equation}
where $Q$ and $I$ are the two Stokes parameters for the emergent radiation in that direction. For practical purpose, $I_l$ and $I_r$ are defined to be the intensity of the light whose electric vector is parallel and perpendicular, respectively, to the plane of scattering. If $I_l=I_r$, the scattered emission would show no polarization. 

For a real atmosphere with some shape, the Stokes parameters of the total radiation emitted in a given line-of-sight direction from the entire surface is found by integrating the slab results over the atmospheric surface, i.e.,
\begin{equation}	\label{eq:qnet}
\begin{split}
Q_{net} &=A_p^{-1}\times\\
&\int_{\scriptstyle projected \hfill \atop 
\scriptstyle surface \hfill}  {d{A_p}\cos 2\psi \left[ {{{Q\left( {\lambda,\mu } \right)} \over {I\left( {\lambda,\mu } \right)}}} \right]\left[ {{{I\left( {\lambda,\mu } \right)} \over F}} \right]F},
\end{split}
\end{equation}
\begin{equation}	\label{eq:inet}
{I_{net}} = A_p^{-1}\times\int_{\scriptstyle projected \hfill \atop 
  \scriptstyle surface \hfill}  {d{A_p}\left[ {{{I\left( {\lambda,\mu } \right)} \over F}} \right]F},
\end{equation}
where $A_p= \pi a c$ and $d A_p$ are the total and differentiated \textit{projected} surface area, respectively, $\psi$ is an angle related to the orientation of the local scattering plane, and $\pi$$F$ is the net flux from the whole surface. The quantities in the square brackets are the plane-parallel results which are tabulated in \cite{chandra} and \cite{shapiro82}.

\begin{figure}
\begin{center}
\includegraphics[width=7.5cm, angle=0]{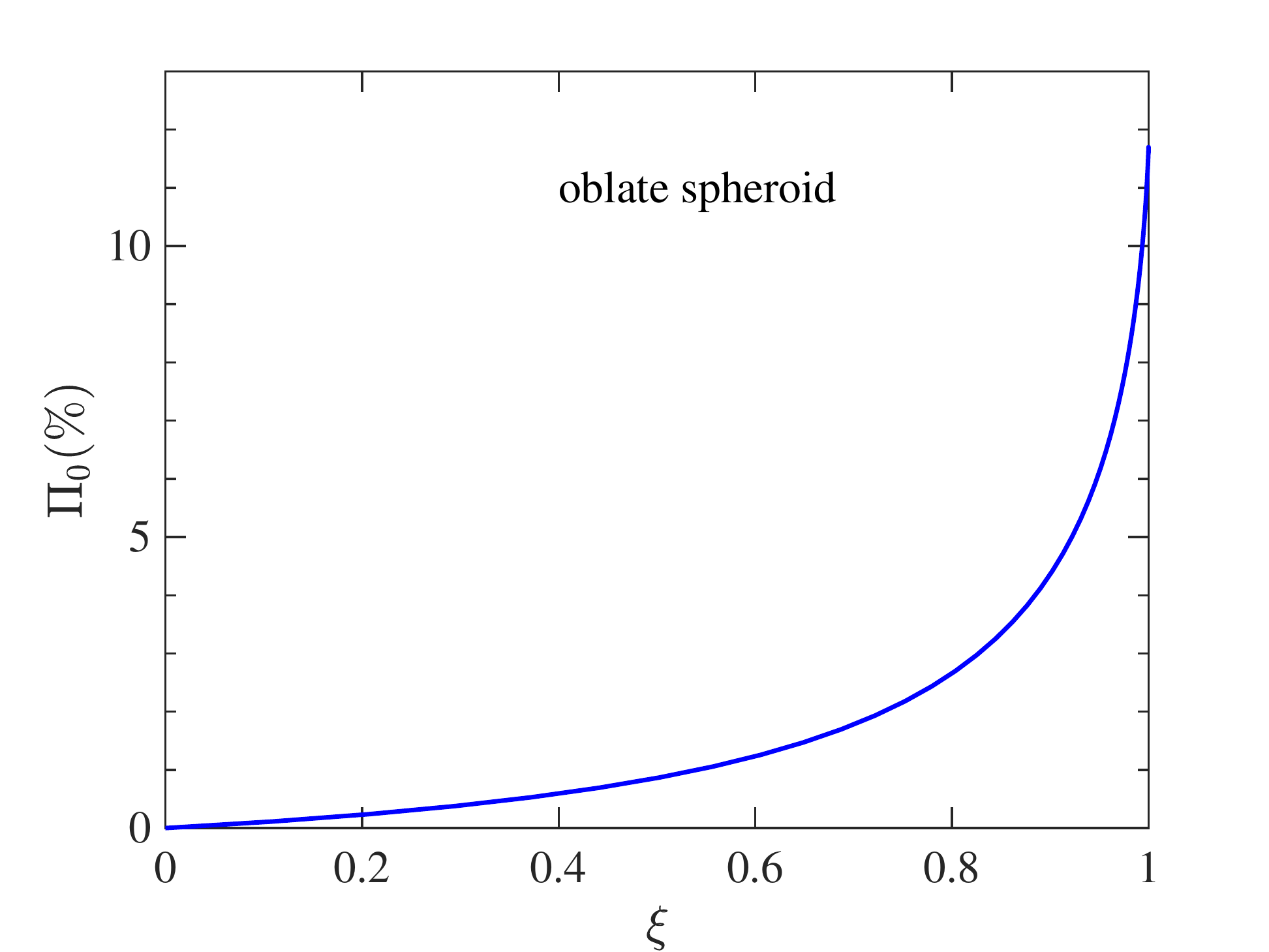} 
\includegraphics[width=7.5cm, angle=0]{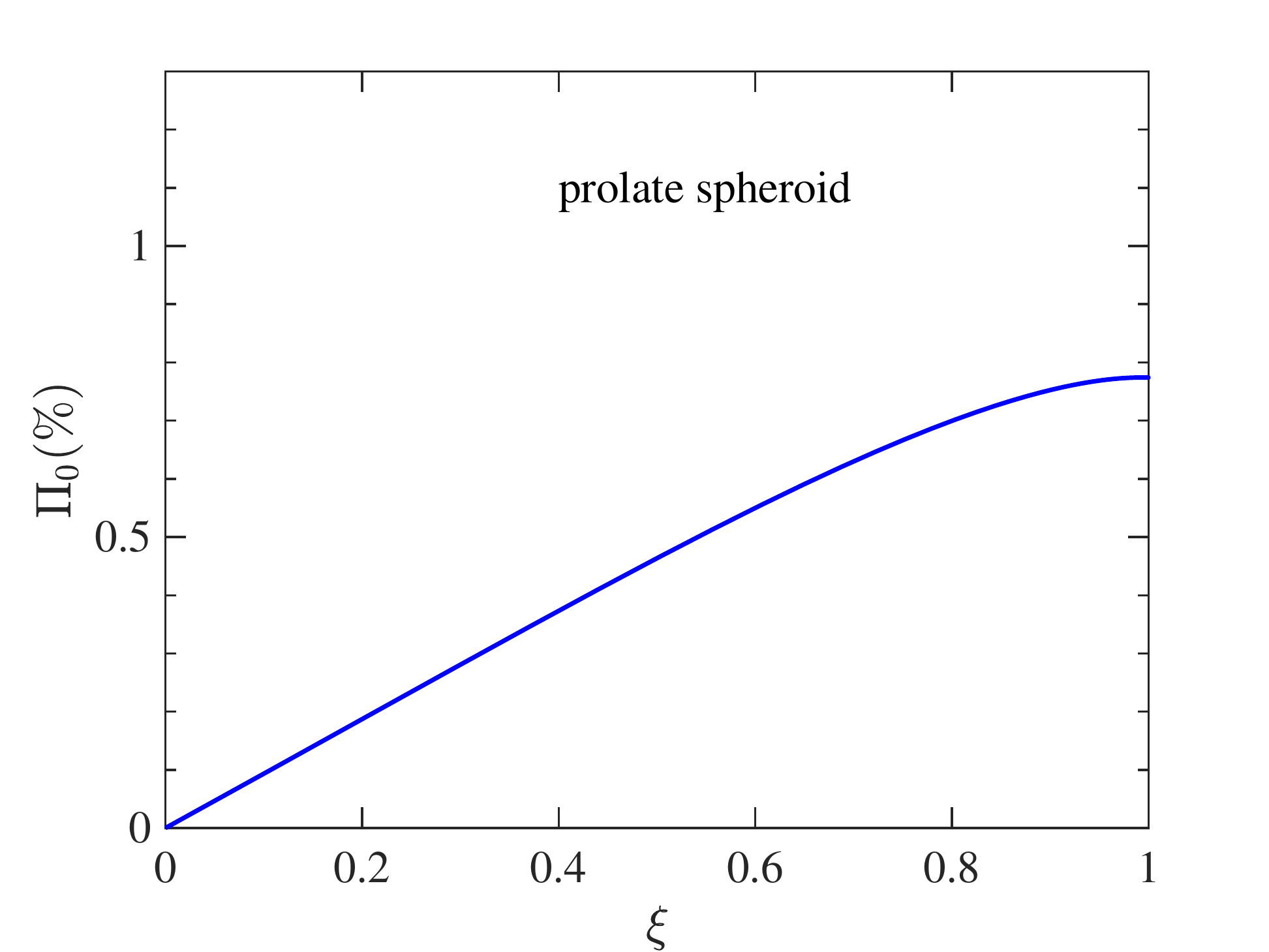}
\caption{The maximum degree of polarization of the emission from oblate and prolate spheroidal atmospheres, respectively, as a function of their asphericity parameters $\xi$, viewed perpendicular to the symmetry axis of the spheroids ($i=90^{\circ}$). }   \label{fig:pola82}
\end{center}
\end{figure}

The asphericity of a spheroidal atmosphere is para-metrized by the parameter
\begin{equation}
\xi  \equiv \left\{ {\begin{array}{*{20}{c}}
1 - (c/a), ~ \mbox{for}~ a > c ~\mbox{(oblate)}\\
1 - (a/c), ~ \mbox{for}~ a < c ~\mbox{(prolate)},
\end{array}} \right.
\end{equation} 
where $a$ is the equatorial radius, and $c$ is the distance along the symmetry axis from pole to center. The surface integrals in equations (\ref{eq:qnet}-\ref{eq:inet}) can be evaluated numerically by quadrature, using a third-order spline fitted to the tabulated values of $[Q/I]$ and $[I/F]$. Then the net polarization degree is
\begin{equation}
\Pi_0= \frac{Q_{net}}{I_{net}},
\end{equation}
which we consider to be the maximum degrees of polarization from an oblate or prolate spheroidal atmosphere with a given asphericity.  $\Pi_0$ calculated this way as a function of $\xi$ are shown in Figure \ref{fig:pola82} (also see Fig. 2 of \cite{shapiro82}). 

Based on the anisotropic distribution of the dynamical ejecta and the disk wind found in simulations \citep{foucart14,kyutoku15,fernandez18}, we will assume that the shapes of the dynamical ejecta and the disk wind are oblate and prolate with $\xi=0.8$ and $0.6$, respectively. We further assume that during the expansion of the ejecta, the photosphere keeps its morphology (i.e., $\xi$ fixed) in time, so $\Pi_0$ does not change.

Then, according to Figure \ref{fig:pola82}, for an observer whose line of sight is in the equatorial plane, the maximum net polarization from the dynamical ejecta and the disk wind is $3\%$ and $0.6\%$, respectively. These values correspond to an early time when the photosphere is right at the free-neutron layer of either ejecta component, where the electron scattering opacity dominates (see below).

In the BH-NS kilonova scenario, the total luminosity is calculated as the sum of those ejecta components (i.e., dynamical ejecta, disk wind, and free neutrons), though at different times it is dominated by different components (see Figure \ref{fig:dyn} below). Similarly, the observed polarization is always contributed from all components. However, integrating the plane-parallel result over the total projected surface area of all components at any time is complicated. Therefore, as an approximation, we let the observed polarization be equal to that of whichever ejecta component that dominates the luminosity at that epoch.


\subsection{Scattering optical depth at the photosphere $\tau_{\rm es}$}		\label{sec:taues}

In the kilonova scenario, the opacity of the bulk of the ejecta (other than the free-neutron layer) shall be absorption dominated, plus some scattering by electrons from partial ionization of r-process products. The optical depth of the hot ejecta is given by $\tau=\int{\kappa \rho dR}$, here $\kappa$ includes the contribution from both scattering and absorption (mainly bound-bound transitions of the r-process elements), and $\tau=1$ at the photosphere $R_{ph}$. However, during the first few hours, the photosphere is located in the outer layers of the ejecta, whose opacity is dominated by electron scattering, thus $\tau_{es} \equiv \tau_{es}(R_{ph}) \simeq \tau(R_{ph})=1$. 
When the photosphere moves into the layers which are mainly composed of r-process nuclei, $\tau_{es} \simeq \kappa_{es} \rho R_{ph}$ decreases with time due to the ever dropping $\kappa_{es}$. 

For the heating sources of kilonova emission, we consider the radioactive decay in the dynamical ejecta and the disk wind, and the $\beta$-decay of the free neutrons. A free neutron layer (fast tail) may exist in the outer layer(s) of dynamical ejecta or disk wind, or both (\S\ref{sec:n-component}). Here, we consider two limiting  cases: free neutrons are contained only in the outermost layer of (1) the dynamical ejecta, or (2) of the disk wind, and calculate the light curves, $\tau_{es}$, and the net polarization. 

The procedure that we take to calculate $\tau_{es}$ is the same as in \cite{matsumoto18}. We take a power-law density profile used in \cite{matsumoto18} for both the dynamical ejecta and the disk wind. Following the merger, the ejecta velocity structure approaches that of a homologous expansion, with the faster matter moving ahead of slower ones \citep{rosswog14}. Thus the layer that is mainly composed of free neutrons would be the fastest tail in the mass-velocity distribution, meanwhile the bulk of the ejecta or is treated as a base component which has a power-law density profile \citep{hotokezaka13, nagakura14} as
\begin{equation}
\begin{split}
{\rho _{base}}\left( {v,t} \right) &= \begin{array}{*{20}{c}}{\frac{{{M_{base}}}}{{4\pi {{\left( {{v_{b}}t} \right)}^3}}}{{\left( {\frac{v}{{{v_{b}}}}} \right)}^{ - {\beta_{b}}}}\left[ {\frac{{{\beta_{b}} - 3}}{{1 - {{\left( {\frac{{{v_{t}}}}{{{v_{b}}}}} \right)}^{3- {\beta_{b}}}}}}} \right],}\end{array} \\
&{{v_{b}} < v < {v_{t}}},
\end{split}
\end{equation}
where $M_{base}$, $v_{b}$ and $\beta_{b}$ are the total mass, the lowest velocity, and the power-law index of the base component, $v_{t}$ is the lowest velocity of the fast tail. We set $v_{b} = 0.1$ c, $v_{t} = 0.3$ c, and $\beta_{b} = 4$, for the fiducial values, and $M_{base} = 0.03 M_\odot$ for the dynamical ejecta and  $M_{base} = 0.02 M_\odot$ for the disk wind, respectively. 

The density profile of the free neutrons (fast tail) is given by the following equation \citep{kyutoku14, hotokezaka18}
\begin{equation}
\begin{split}
{\rho _{tail}}\left( {v,t} \right) &= \begin{array}{*{20}{c}}{\frac{{{M_n}}}{{4\pi {{\left( {{v_n}t} \right)}^3}}}{{\left( {\frac{v}{{{v_n}}}} \right)}^{ - {\beta_{t}}}}\left[ {\frac{{{\beta_{t}} - 3}}{{1 - {{\left( {\frac{c}{{{v_n}}}} \right)}^{3-{\beta_{t}}}}}}} \right],}\end{array}\\ 
&{{v_{t}} < v},
\end{split}
\end{equation}
where $M_{n}=10^{-4}{M_\odot }$, $v_{n}=0.5$ c,  and $\beta_{t}=6$ are the total mass and the lowest velocity of the free neutrons, and the power-law index of the fast tail's profile, respectively. The ejecta structure is composed of a series of $v$-shells. For any one shell, the exterior mass is 
\begin{equation}
M\left( { > v} \right) = \int_{vt}^{ct} {dr4\pi {r^2}\left( {{\rho _{base}} + {\rho _{tail}}} \right)}. 
\end{equation}

In order to calculate $\tau_{es}$ and the light curve of a free-neutron precursor, chemical composition of all the mass shells are needed. Note that for one of the two limiting cases, free neutrons exist only in the fast tail of dynamical ejecta, then the whole disk wind consists of r-process materials only, i.e., $\chi_r(v,t) = 1$, and vice versa. We assume that the fastest ($v>v_n$) layer of the ejecta contain free neutrons only. For the inner part of the high-speed tail component (i.e., $v_{t}<v<v_n$), the mass fractions of the r-process elements, free neutrons, and protons (the product of the neutrons' $\beta$-decay), respectively, are given by
\newline 
\begin{equation}
{\chi _r}\left( {v,t} \right) = \frac{2}{\pi }\arctan \left[ {{{\left( {\frac{{M\left( { > v} \right)}}{{{M_n}}}} \right)}^\alpha }} \right],
\end{equation}
\begin{equation}
{\chi _n}\left( {v,t} \right) = \left( {1 - {\chi _r}} \right){e^{ - t/{t_n}}}\\,
\end{equation}
\begin{equation}
{\chi _p}\left( {v,t} \right) = \left( {1 - {\chi _r}} \right)\left( {1 - {e^{ - t/{t_n}}}} \right),
\end{equation}
\newline
where the index $\alpha$ describes how sharply the nucleon (neutrons, in this case) abundance decreases toward inner shells with $v < v_n$ \citep{matsumoto18} . 
Here we set $\alpha=10$, while other values do not change the results significantly. $t_n$ = 900 s is the neutrons' $\beta$-decay timescale. For slow ejecta shells with $v<v_{t}$, we set $\chi_r=1$.

Photos escape the ejecta by diffusion. For each mass shell, we can calculate its photon diffusion time (Eq. \ref{eq:tpeak}). Conversely, at any time t we can locate the `diffusion shell' whose photon diffusion time is $t_{diff} = t$ by solving the equation of $\tau = c/v$. Hence we can get the velocity of the diffusion shell $v_d$. The optical depth is given by
\begin{equation}
\tau \left( {v,t} \right) = \int_{vt}^{ct} {dr\left[ {{\kappa _r}{\rho _{base}} + \left( {{\kappa _r}{\chi _r} + {\kappa _{es,H}}{\chi _p}} \right){\rho _{tail}}} \right]} 
\end{equation}
where we set $\kappa_{r,}{_{dyn}} = 10$ cm$^2$/g and $\kappa_{r,}{_{wind}} = 1$ cm$^2$/g for the bound-bound opacity of the r-process elements in the dynamical ejecta and the disk wind, respectively \citep{kasen17}. In Eq. (13), we ignore the contribution to $\kappa_{es}$ by electrons supplied by the ionized of the r-process elements due to its small value. 

The bolometric luminosity is calculated following the Arnett model (\cite{arnett82}; see also \cite{metzger17}) by considering the total mass $M(> v_{d})$ exterior to the diffusion shell. The mass of the r-process elements and free neutrons are evaluated, respectively, by
\begin{equation}
{M_r}\left( { > {v_{d}}} \right) = \int_{{v_{d}}t}^{ct} {dr4\pi {r^2}\left( {{\rho _{base}} + {\chi _r}{\rho _{tail}}} \right)}, 
\end{equation}
\begin{equation}
{M_n}\left( { > {v_{d}}} \right) = \int_{{v_{d}}t}^{ct} {dr4\pi {r^2}} {\chi _n}{\rho _{tail}}.
\end{equation}
\newline 
Then the luminosity is
\begin{equation}	\label{eq:Lt}
L(t) \simeq {q_r}{M_r}\left( { > {v_{d}}} \right) + {q_n}{M_n}\left( { > {v_{d}}} \right)
\end{equation}
\newline 
where the specific heating rates due to the radioactive decay of the r-process elements and free neutrons are given by \cite{wanajo14} and \cite{kulkarni05} respectively as follows,
\begin{equation}	\label{eq:qr}
{q_r} = 2 \times {10^{10}}{\left( {\frac{t}{{day}}} \right)^{ - 1.3}}erg{s^{ - 1}}{g^{ - 1}},
\end{equation}
\begin{equation}
{q_n} = 3 \times {10^{10}}erg{s^{ - 1}}{g^{ - 1}}.
\end{equation}
\newline
The photospheric effective temperature is given by $T_{ph} = (L/4π\sigma_{SB}R_{ph}^2 )^{1/4}$, where $\sigma_{SB}$ is the Stefan-Boltzmann constant.

At the photosphere $R_{ph }= v_{ph}t$, we evaluate $\tau_{es}$ as 
\begin{equation}
{\tau _{es}} = \int_{v_{ph}t}^{ct} {dr{\kappa _{es,H}}\left[ {\frac{x}{A}\left( {{\rho _{base}} + {\chi _r}{\rho _{tail}}} \right) + {\chi _p}{\rho _{tail}}} \right]}, 
\end{equation}
where the contributions by electrons from the ionization of the r-process elements are included (the first and second terms). Following \cite{matsumoto18}, we adopt the mean mass number of the r-process elements $A = 80$, and set the degree of ionization of the r-process nuclei to be $x=1$. 

\begin{figure}
\begin{center}
\includegraphics[width=7.5cm, angle=0]{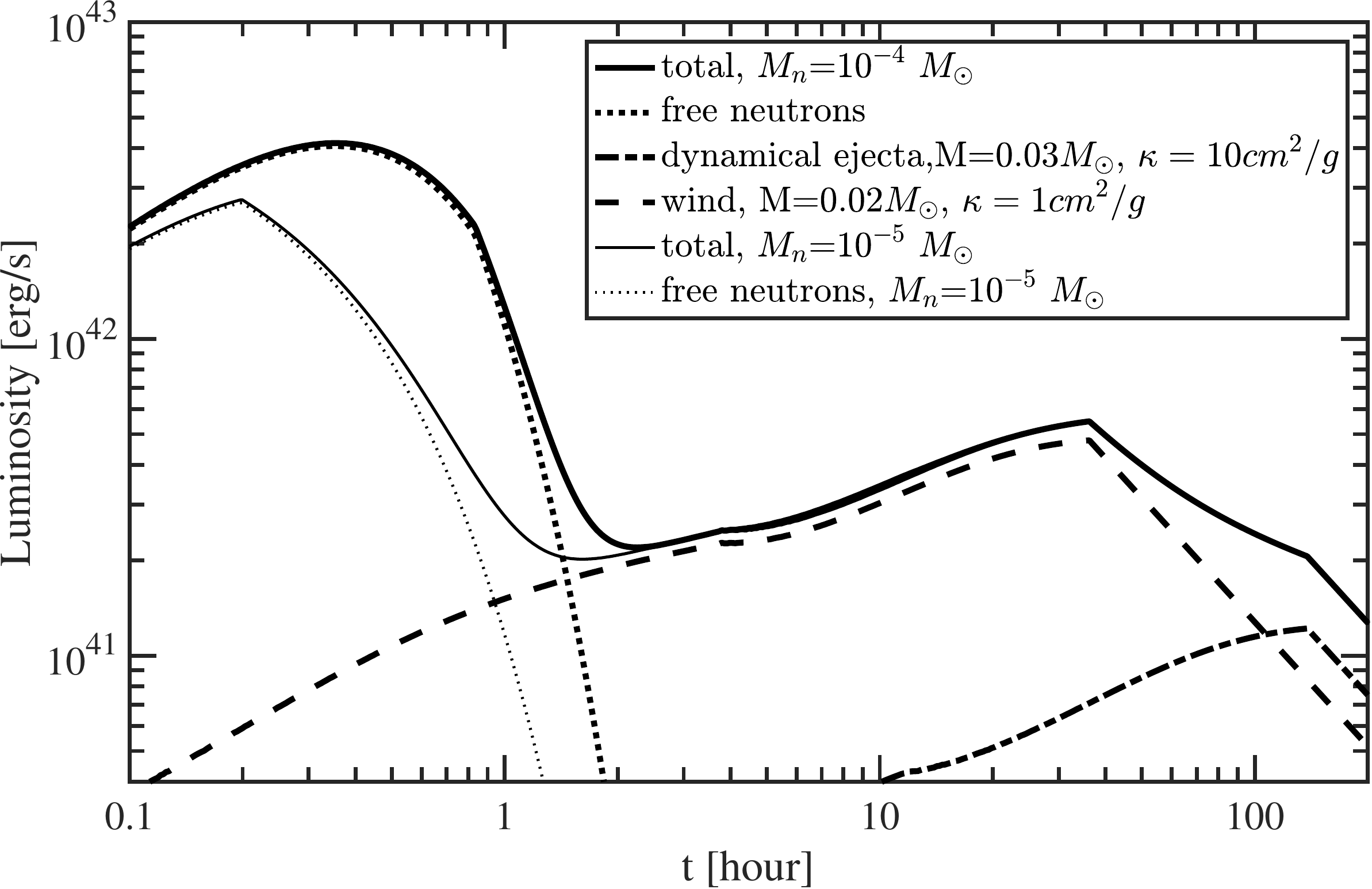}
\includegraphics[width=7.5cm, angle=0]{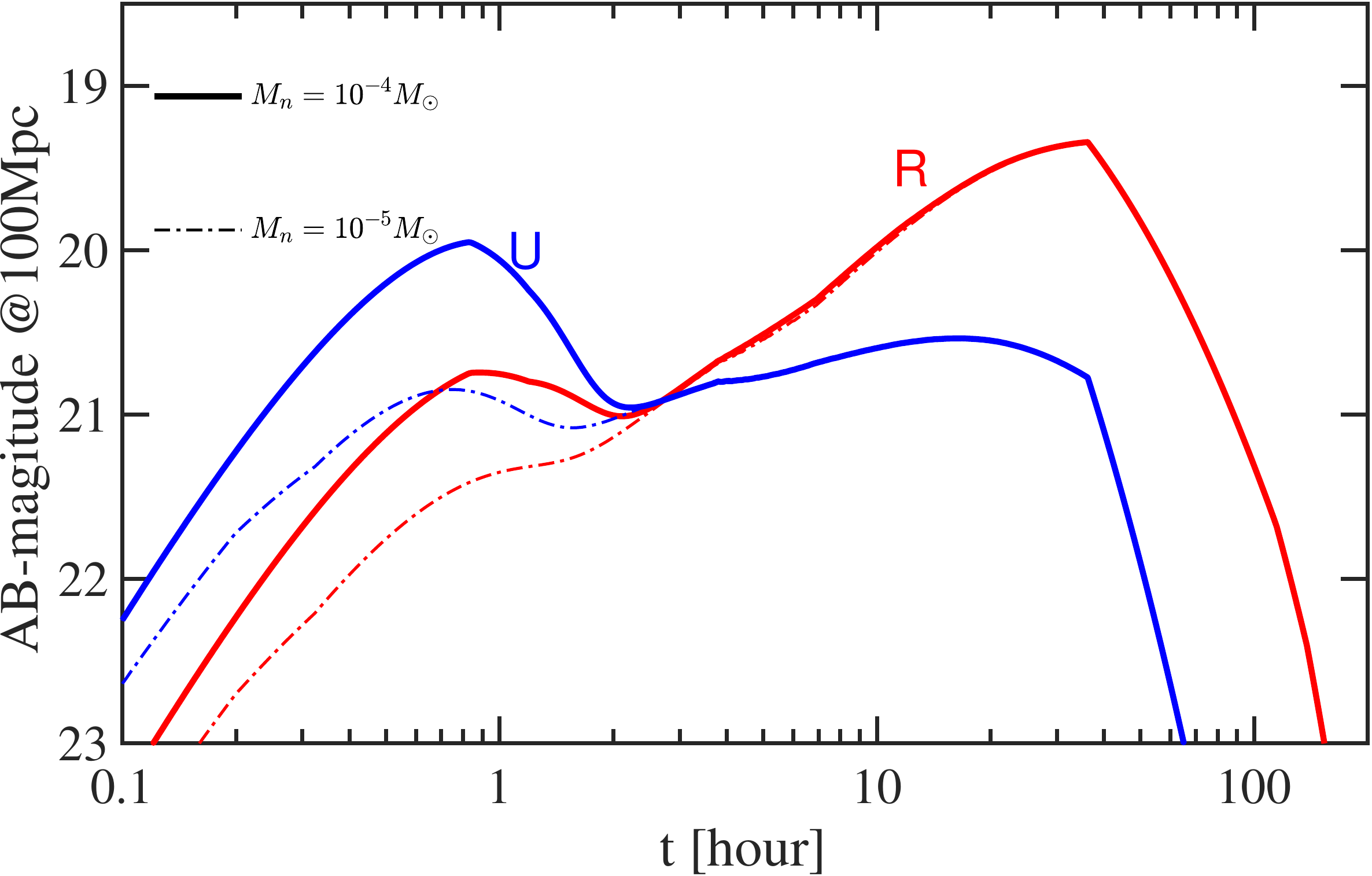}
\caption{
The bolometric luminosity light curve (top) and the light curves in the $U$ and $R$ bands (bottom), respectively, for a kilonova from a BH-NS merger, assuming free neutrons survived in the outmost layers of dynamical ejecta. The $\beta$-decay of those free neutrons causes a precursor at $t\sim 1$ hr. Here, two values of free neutron layers mass are considered: $10^{-5}$ and $10^{-4}$ $M_{\odot}$.}		\label{fig:dyn}
\end{center}
\end{figure}

\begin{figure}
\begin{center}
\includegraphics[width=7.5cm, angle=0]{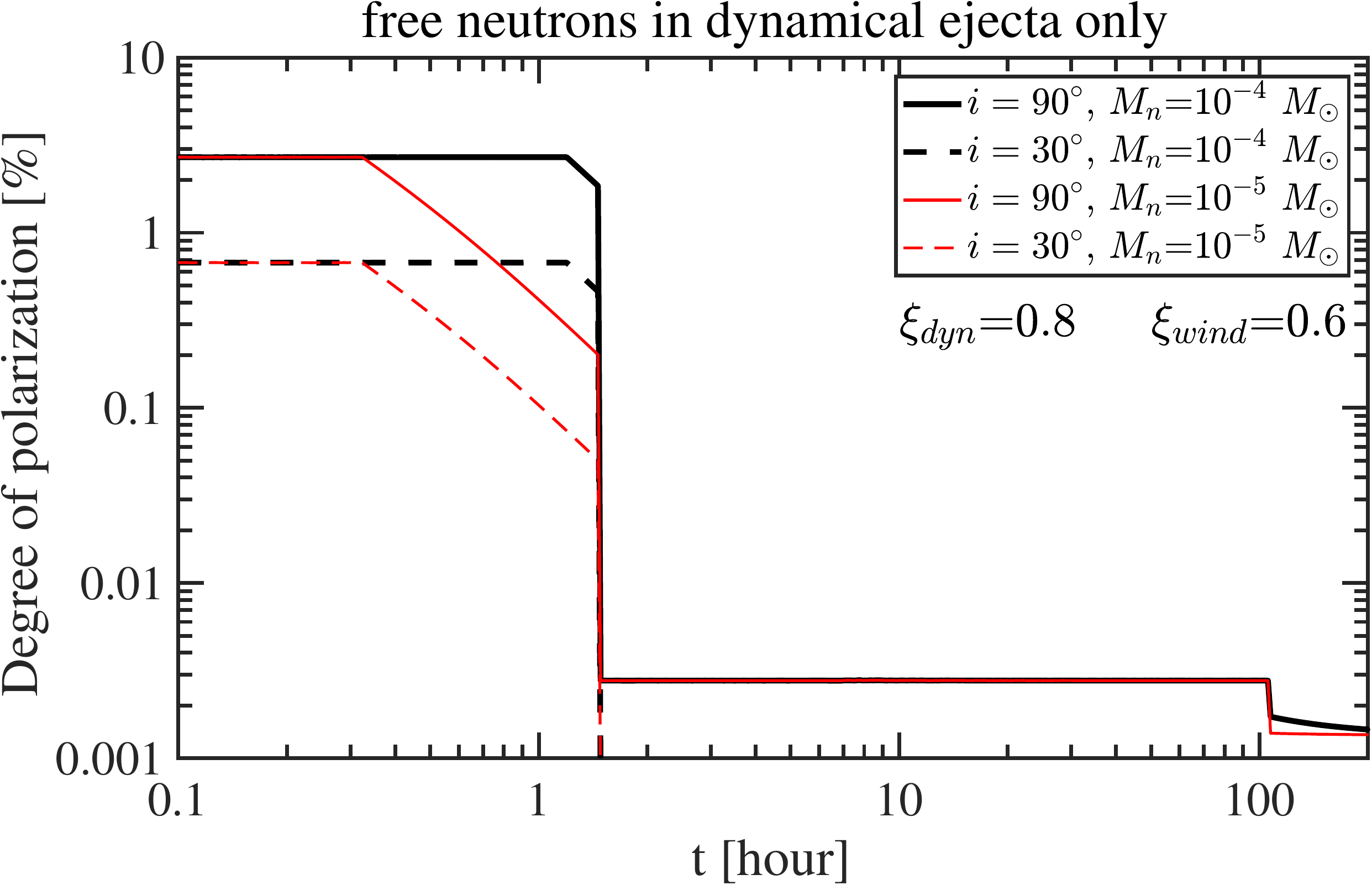}
\includegraphics[width=7.5cm, angle=0]{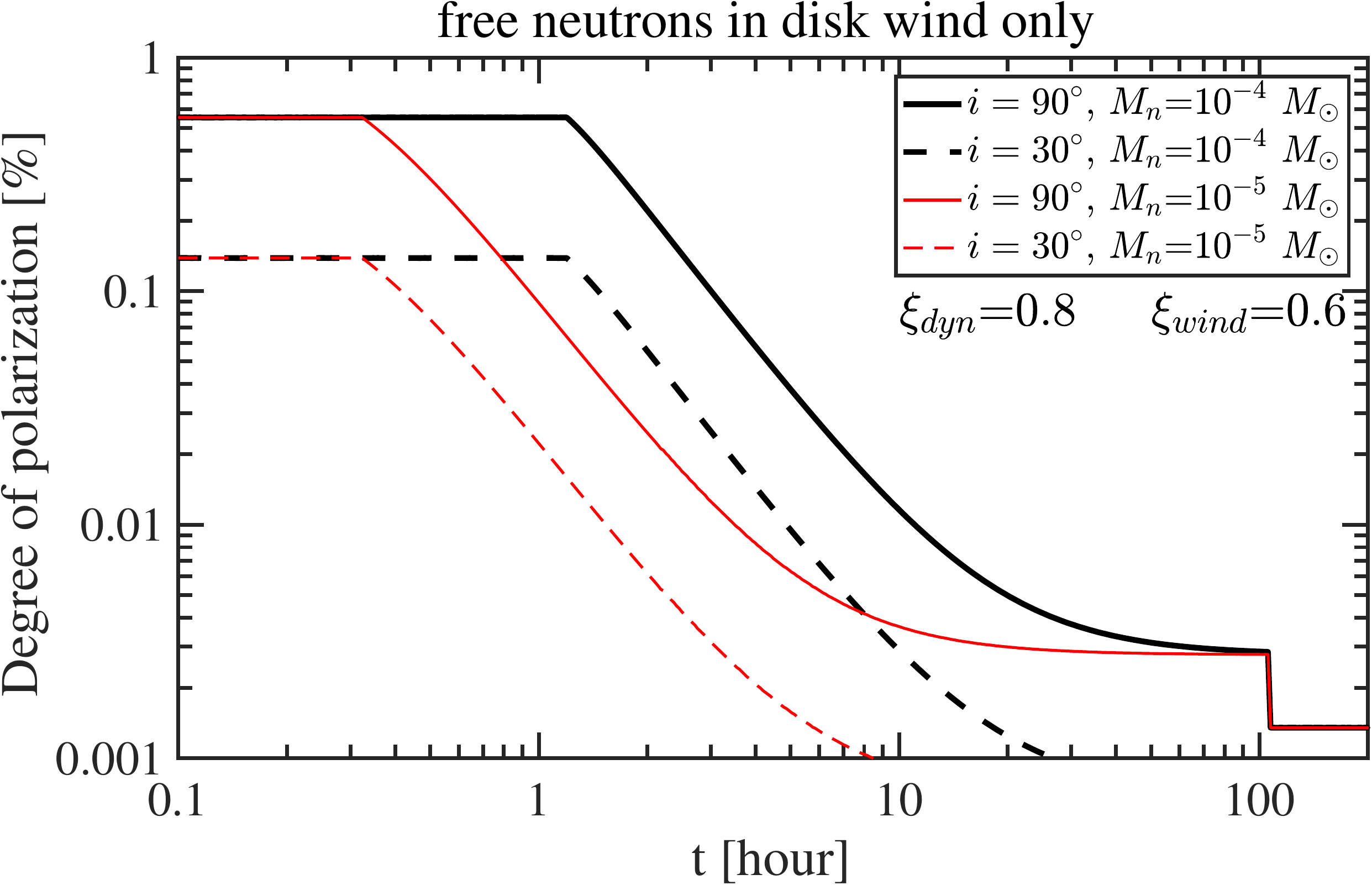}
\caption{The degrees of polarization of the observed emission for the case in which the fast tail of the dynamical ejecta contains free neutrons (\textit{top}), and for the case in which the fast tail of the disk wind contains free neutrons (\textit{bottom}). }		\label{fig:wind}
\end{center}
\end{figure}

\begin{figure}
\begin{center}
\includegraphics[width=7.5cm, angle=0]{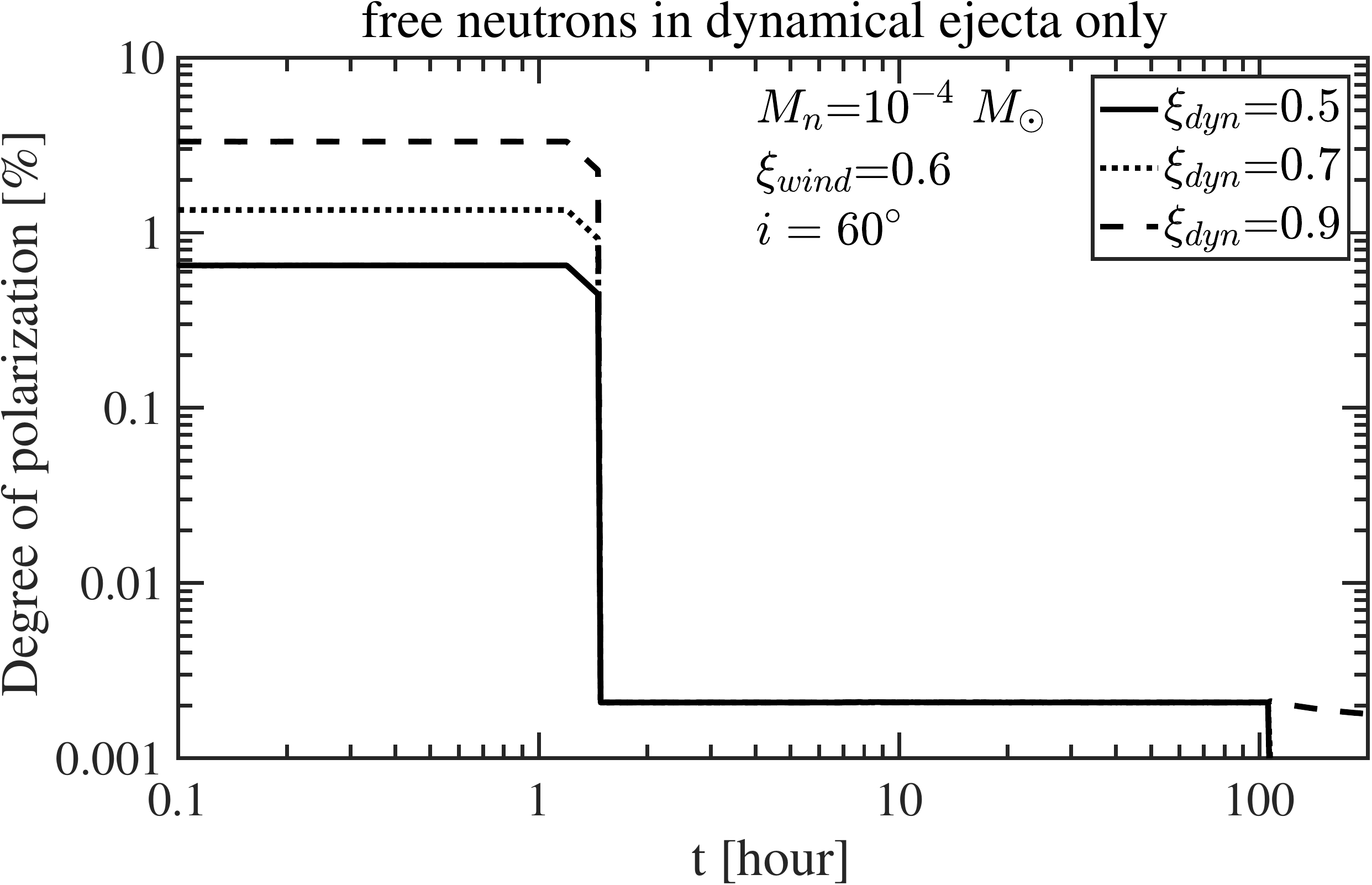}

\includegraphics[width=7.5cm, angle=0]{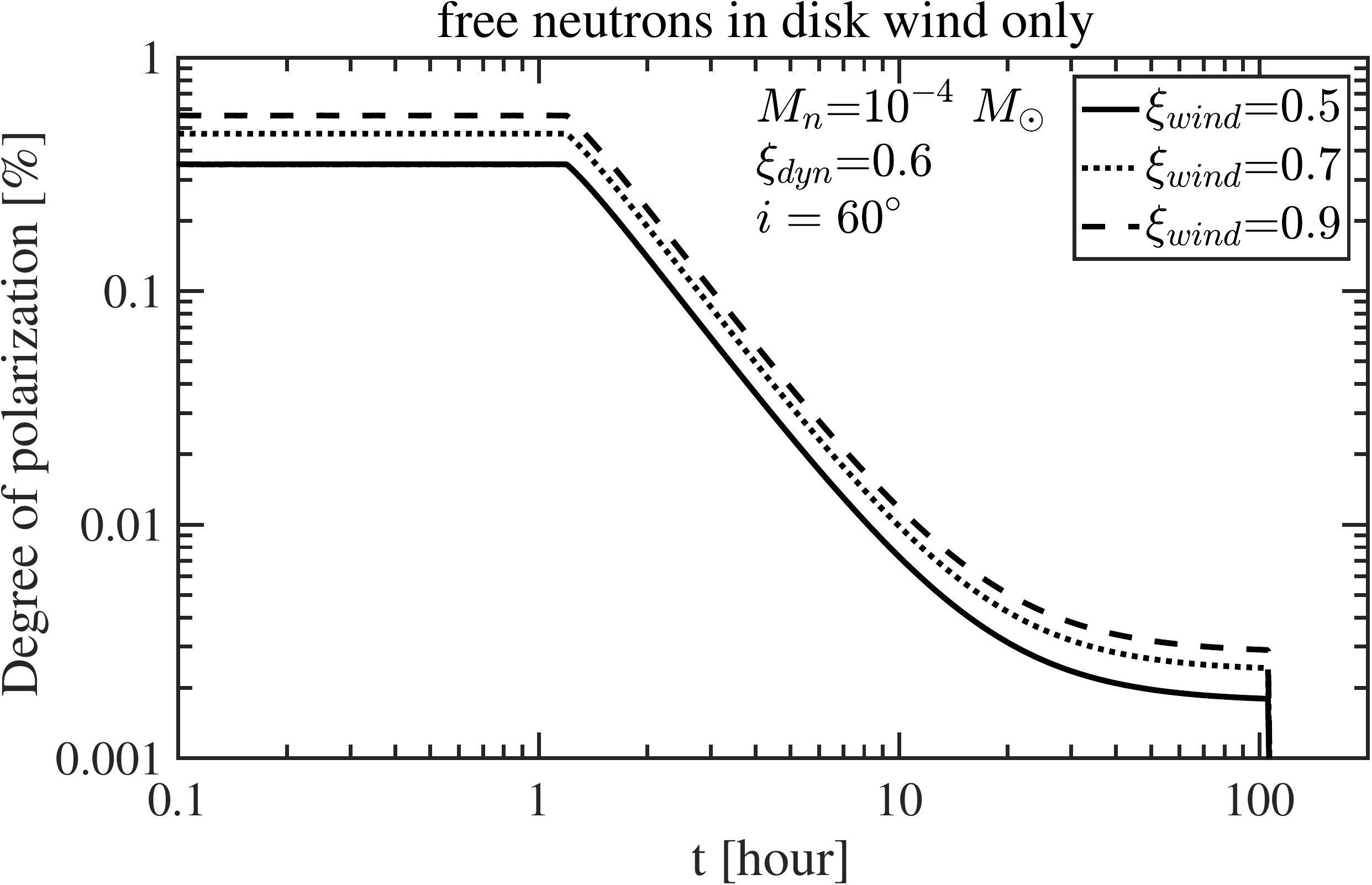}
\caption{Same as in Fig. \ref{fig:wind} but with varying asphericity parameters for the dynamical ejecta (\textit{top}) and the disk wind (\textit{bottom}), respectively.}	\label{fig:xi}
\end{center}
\end{figure}

\section{Results}

With equation (\ref{eq:Lt}) and $T_{ph}$, the time evolution of the bolometric luminosity and the $U$ and $R$ bands apparent magnitudes are calculated and shown in Figure \ref{fig:dyn}. It shows that, the free-neutron heating causes a bump in the light curve at $\sim1$ hr (the diffusion time of the free-neutron layer) after the merger, a feature already predicted in \cite{metzger15} and \cite{matsumoto18}. 

The degrees of polarization as a function of time are calculated according to equation (\ref{eq:pinet}) and are shown in Figure \ref{fig:wind}. The thick solid and dashed curves correspond to viewing angles of $90^{\circ}$ and $30^{\circ}$, respectively (Eq. 2). Here we emphasize that, in any epoch, the degree of polarization is not the sum of those of the two base components, but is determined by whichever component that dominates the luminosity.
 
For the case where free neutrons are contained only in the dynamical ejecta (Figure \ref{fig:wind}, \textit{top}), initially the degree of polarization is almost constant with $\Pi_{\rm net}$ $\simeq$ $\Pi_0$ until an abrupt decrease at $\sim1$ hr after the merger. The degree of polarization can be up to $3\%$ depending on the viewing angle. For the other case (free neutrons in the disk wind only; Figure \ref{fig:wind}, \textit{bottom}), the degree of polarization shows a similar plateau during the first $\sim1$ hr, with a lower $\Pi_{\rm net}$ up to $0.6\%$, and then smoothly decreases afterwards. 

These temporal behaviors of $\Pi_{\rm net}$ can be understood as follows. Initially, the photosphere is located in the free neutron layer and $\tau_{es}\simeq1$ resulting in the polarization $\Pi_{\rm net}$ $\simeq$ $\Pi_0$. In the case that free neutrons exist in the dynamical ejecta, the shape of the  photosphere is that of the dynamical ejecta, i.e., an oblate spheroid (Figure \ref{fig:comp}) whose maximum polarization is $\approx 3\%$ for $\xi = 0.8$ according to Figure \ref{fig:pola82}. On the other hand, if free neutrons exist in the outermost layer of the disk wind, whose shape is a prolate spheroid, the maximum  polarization is $\approx 0.6\%$ for $\xi =0.6$. 

As the photosphere recedes to the r-process element-rich ejecta ($v < v_n$), $\tau_{es}$ drops quickly, resulting in the significant decrease of the polarization degree. Because the observed degree of polarization is determined by whichever of the two base components that dominates the luminosity, there is a difference in the trend for the two scenarios. Fig. \ref{fig:wind} (top) shows an abrupt drop, whereas in Fig. \ref{fig:wind} (bottom) the drop is smoother. In the former case that free neutrons are contained in the dynamical ejecta, the disk wind component starts to dominate the luminosity after $\sim1$ hr since the merger (see Figure \ref{fig:dyn}), whose $\tau_{es}$ is just related to the ionization of the r-process nuclei. Due to the constant quantities $x=1$ and A=$80$ we set, the degree of polarization is constant and small ($\sim0.01\%$) at that epoch. In the other case, the disk wind contains free neutrons and also dominates the later luminosity, thus there is a smoother decrease. The late-time polarization in our result is consistent with that of \cite{bulla18} though we use a different method.

The dependence of polarization degree on the asphericities of dynamical ejecta and disk wind is shown in Figure \ref{fig:xi}. As was shown in Figure \ref{fig:pola82}, the polarization degree increases with the asphericity. In other words, the early $\sim 1$ hr observation of polarization degree can provide a constraint on the morphology of the dynamical ejecta or the disk wind. 

There are a few discontinuities in the time evolution of the polarization degree shown in Figures \ref{fig:wind}-\ref{fig:xi}. Under our prescription for the polarization degree in the presence of multiple ejecta components (\S \ref{sec:pi0}), these are due to the changes of dominance of the total emission among these components. Note that those abrupt changes of polarization degree also correspond to changes of the polarization angle. The polarization angle is defined as the angle at which the transmitted intensity is at maximum \citep{kasen03}. The polarization angle of a spheroid (oblate or prolate) is perpendicular to its own long axis. Thus since the polarization angles for dynamical ejecta (oblate spheroid) and disk wind (prolate spheroid) are perpendicular, we expect that the observed polarization angle will experience a noticeable change, for instance, at $\sim1$ hr.

\section{Summary and Discussion}

So far gravitational wave observation found only one NS-NS merger event, with its multiwavelength EM counterparts including a kilonova. BH-NS mergers have been predicted to occur with a rate similar to NS-NS mergers and to produce a kilonova signature as well, but they still remain to be detected. Because the ejecta components from a BH-NS merger are predicted to be highly anisotropic, therefore any possible existence of free electrons in the atmosphere of those ejecta components could cause a non-zero polarization. As was found in \cite{matsumoto18}, the degree of polarization strongly relies on the existence of a free neutrons in the fastest component of the ejecta. While previous works \citep{bulla18,matsumoto18} concentrated to the context of NS-NS mergers, in this paper we focus on the BH-NS mergers, a case potentially more susceptible to a polarization signature. 

Assuming the dynamical ejecta and disk wind  components have masses of  $0.03M_{\odot}$ and $0.02M_{\odot}$, and are in shapes of oblate and prolate spheroids, respectively, we estimate the degree of polarization at the first $\sim 1$ hr can be up to 3\% and 0.6\% if free neutrons ($10^{-4} M_{\odot}$) survived in the fastest component of the dynamical ejecta or the disk wind, respectively. The degree of polarization drops afterwards when the free neutron layer becomes transparent. For the case that free neutrons only exist in the dynamical ejecta, the  degree of polarization shows an abrupt drop to a negligible level. The decline of the polarization degree is much smoother if free neutrons exist only in the disk wind. 

Future detection of polarization at $\sim1$ hr can determine the distribution of free neutrons and the morphology of dynamical ejecta and disk wind. Specifically, if the polarimetric observation is performed at the early timescale $t<\sim1$ hr, we may detect the maximum polarization degree of kilonova emission and its evolution at later times. (1) If a $\Pi_{\rm net}\gtrsim 0.6\%$ is detected, it would mean the polarization is dominated by the free neutrons in the dynamical ejecta. Combining with a viewing angle determined from other channels, one could determine the asphericity of the dynamical ejecta. (2) If $\Pi_{\rm net}$ is found to be $\lesssim 0.6\%$, it would not be certain as to which one of the base components dominates the polarization, because both components can produce such a relatively low degree. However, it can be constrained by later-time evolution of polarization. For instance, if an abrupt (smooth) decline of the polarization degree follows, it would imply that the free neutrons only exist in the outer layer of dynamical ejecta (disk wind). 

The BH-NS dynamical ejecta are found to occupy only an azimuthal angle of $\sim \pi$ \citep{foucart14,Kawaguchi15,kyutoku15,kyutoku18}, whereas so far we assumed the observer is on the approaching side of the dynamical ejecta (Figure \ref{fig:comp}). If the dynamical ejecta is receding with respect to the observer, our results are unchanged for the case where the free neutrons reside in the disk wind only (Figures \ref{fig:wind}-\ref{fig:xi}, \textit{bottom}). However for the other case where only the dynamical ejecta contains free neutrons, an observer on the receding side would probably not see a free-neutron precursor because it is likely blocked by slower ejecta or the disk wind. In that case the early plateau of the polarization in Figures \ref{fig:wind}-\ref{fig:xi} (\textit{top}) might disappear or be reduced depending on the azimuthal offset of the line of sight. 

In calculating the dynamical ejecta's contribution to the overall light curve, we used the simple spherically symmetric Arnett model since we focus on the polarization property. A more careful calculation shall take into account the highly anisotropic geometry of this component, such as were done in \cite{kawaguchi16} and \cite{huang18}. According to these authors, most of photons diffuse vertically toward the latitudinal surface rather than to the radial edge due to a large contrast between areas of the two surfaces. This would cause the dynamical ejecta's light curve to peak earlier than what was predicted in the spherical approximation. The latter can be given by equation (\ref{eq:tpeak}) where the parameters shall take values appropriate for the base dynamical ejecta component (\S \ref{sec:taues}). Generically this time corresponds to when the deepest part of the ejecta is visible. Keeping other conditions (mass, speeds and opacity) same as what we adopted, the latitudinal diffusion effect shortens this time by a factor of $\approx [2\pi \theta_{\rm ej} v_b/(\psi_{\rm ej} v_t)]^{-1/2} \approx 3$, where $\theta_{\rm ej} \simeq 1/5$ and $\psi_{\rm ej} \simeq \pi$ are the latitudinal and azimuthal opening angles of the ejecta, respectively (cf. equation 10 of \cite{kawaguchi16}). Correspondingly, the peak luminosity from this ejecta component would increase by this factor to the power of 1.3 (cf. equations \ref{eq:Lt}-\ref{eq:qr}). Therefore this effect might cause a slow decline or an excess in the late light curve, but it hardly changes the light curve earlier than $t\approx$ 40 hrs presented in Figure \ref{fig:dyn} since the disk wind component still dominates until then. The free-neutron precursor will not be affected either because they are at the radial edge of the ejecta.

Overall, matching our model predictions with future observations will help to shed new lights on understanding the composition of ejecta from a BH-NS merger and the mechanism of kilonova emission.

\section{Acknowledgements}

We thank the anonymous referee for useful comments and suggestions that improved the quality of the paper. This work is supported by NSFC grant 11673078.


\end{CJK*}

\begin{thebibliography}{99}
\bibitem[Arnett(1982)]{arnett82} Arnett, W. D. 1982, ApJ, 253, 785

\bibitem[Bauswein et al.(2013)]{bauswein13} Bauswein, A., Goriely, S., Janka, H.T. 2013, ApJ, 773, 78

\bibitem[Bhattacharya, Kumar \& Smoot(2018)]{bhatta18} Bhattacharya, M., Kumar, P., Smoot, G. 2018, arXiv:1809.00006  

\bibitem[Brown \& McLean(1977)]{brown77} Brown, J. C. \& McLean, I. S. 1977, A\&A, 57, 141 

\bibitem[Bulla et al.(2018)]{bulla18} Bulla, M., Covino, S., Kyutoku, K., et al. 2018, Nature Astronomy, 3, 99 

\bibitem[Chandrasekhar(1960)]{chandra} Chandrasekhar, S. 1960, Radiative Transfer (New York: Dover)


\bibitem[Collins(1970)]{collins70} Collins, G. W. II, 1970, ApJ, 159, 583

\bibitem[Covino et al.(2017)]{covino17} Covino, S., Wiersema, K., Fan, Y. Z., et al., 2017, Nature. Astronomy., 1, 791 

\bibitem[Daniel(1980)]{daniel80} Daniel, J. Y. 1980, A\&A, 86, 198  

\bibitem[Fern\'{a}ndez \& Metzger(2016)]{fernandez16} Fern\'{a}ndez, R., Metzger, B. D. 2016, Annu. Rev. Nucl. Part. Sci., 66, 23

\bibitem[Fern\'{a}ndez et al.(2018)]{fernandez18} Fern\'{a}ndez, R., Tchekhovskoy, A., Quataert, E., Foucart, F., Kasen, D. 2018, MNRAS, 482, 3373

\bibitem[Foucart et al.(2014)]{foucart14} Foucart, F., Deaton, M. B., Duez, M. D., et al. 2014, Phy. Rev. D, 90, 024026



\bibitem[H\"{o}flich(1991)]{hoflich91} H\"{o}flich, P. 1991, A\&A, 246, 481

\bibitem[Hotokezaka et al.(2013)]{hotokezaka13}Hotokezaka, K., Kiuchi, K., Kyutoku, K., Okawa, H., Sekiguchi, Y.i., Shibata, M., Taniguchi, K. 2013, Phys. Rev. D, 87, 024001

\bibitem[Hotokezaka et al.(2018)]{hotokezaka18} Hotokezaka, K.; Kiuchi, K., Shibata, M., Nakar, E., Piran, T. 2018, ApJ, 867, 95

\bibitem[Huang et al.(2018)]{huang18} Huang, Z. Q., Liu, L. D., Wang, X. Y., Dai Z. G. 2018, ApJ, 867, 1

\bibitem[Ishii et al.(2018)]{ishii18} Ishii, A., Shigeyama, T., Tanaka M. 2018, ApJ, 861, 25

\bibitem[Just et al.(2015)]{just15} Just, O., et al., 2015, MNRAS, 448, 541

\bibitem[Karp et al.(1977)]{karp77} Karp, A. H., Lasher, G., Chan, K. L., Salpeter, E. E. 1977, ApJ, 214, 161

\bibitem[Kasen et al.(2003)]{kasen03} Kasen, D., Nugent, P., Wang, L., et al. 2003, ApJ, 593, 788

\bibitem[Kasen et al.(2013)]{kasen13} Kasen, D., Badnell, N. R., Barnes, J. 2013, ApJ, 774, 25

\bibitem[Kasen et al.(2017)]{kasen17} Kasen, D.; Metzger, B.D., Barnes, J., Quataert, E., Ramirez-Ruiz, E. 2017, Nature, 551, 80 

\bibitem[Kasen, Fern\'andez \& Metzger(2015)]{kasen15} Kasen, D., Fern\'andez, R., \& Metzger, B. D. 2015, MNRAS, 450, 1777   

\bibitem[Kawaguchi et al.(2015)]{Kawaguchi15} Kawaguchi, K., Kyutoku, K., Nakano, H., Okawa, H., Shibata, M., Taniguchi, K. 2015, Phys. Rev. D, 92, 024014 

\bibitem[Kawaguchi et al.(2016)]{kawaguchi16} Kawaguchi, K., Kyutoku, K., Shibata, M., Tanaka, M. 2016, ApJ, 825, 1

\bibitem[Kiuchi et al.(2015)]{kiuchi15} Kiuchi, K., Sekiguchi, Y., Kyutoku, K., Shibata, M., Taniguchi, K., Wada, T. 2015, Phys. Rev. D, 92, 064034

\bibitem[Kulkarni (2005)]{kulkarni05} Kulkarni, S. R. 2005, arXiv:astro-ph/0510256

\bibitem[Kyutoku et al.(2013)]{kyutoku13}Kyutoku, K., Ioka, K., Shibata, M. 2013, Phys. Rev. D, 88, 041503

\bibitem[Kyutoku et al.(2014)]{kyutoku14} Kyutoku, K., Ioka, K., Shibata, M. 2014, MNRAS, 437, L6

\bibitem[Kyutoku et al.(2015)]{kyutoku15} Kyutoku, K., Ioka, K., Okawa, H., Shibata, M.;, Taniguchi, K. 2015, Phys. Rev. D, 92, 044028

\bibitem[Kyutoku et al.(2018)]{kyutoku18} Kyutoku, K., Kiuchi, K., Sekiguchi, Y., Shibata, M., Taniguchi, K. 2018, Phys. Rev. D, 97, 023009

\bibitem[Li \& Paczy\'{n}ski (1998)]{li98} Li, L. X., Paczy\'{n}ski, B. 1998, ApJL, 507, 59

\bibitem[Matsumoto(2018)]{matsumoto18} Matsumoto, T. 2018, MNRAS, 481, 1008

\bibitem[Metzger et al.(2015)]{metzger15}  Metzger, B. D., Bauswein, A., Goriely, S., Kasen, D. 2015, MNRAS, 446, 1115
 
\bibitem[Metzger(2017)]{metzger17} Metzger, B. D. 2017, Living Reviews in Relativity, 20, 3

\bibitem[Nagakura et al.(2014)]{nagakura14} Nagakura, H., Hotokezaka, K., Sekiguchi, Y., Shibata, M., Ioka, K. 2014, ApJ, 784, L28

\bibitem[Oechslin \& Janka (2006)]{oechslin06} Oechslin, R., Janka, H. T. 2006, MNRAS, 368, 1489

\bibitem[Radice et al.(2018a)]{radice18} Radice, D., Perego, A., Hotokezaka, K., et al. 2018, ApJ, 869, 130 

\bibitem[Radice et al.(2018b)]{radice18b} Radice, D., Perego, A., Hotokezaka, K., et al. 2018, ApJL, 869, L35  

\bibitem[Rosswog et al.(2014)]{rosswog14}Rosswog, S., Korobkin, O., Arcones, A., Thielemann, F. K., Piran, T. 2014, MNRAS, 439, 744

\bibitem[Rybicki \& Lightman(1979)]{rybicki} Rybicki, G. B. \& Lightman, A. P., Radiative Processes in Astrophysics, Wiley-VCH

\bibitem[Shapiro \& Sutherland(1982)]{shapiro82}  Shapiro, P. R. \& Sutherland, P. G. 1982, ApJ, 263, 902

\bibitem[Siegel \& Metzger(2017)]{siegel17} Siegel, D. M., Metzger, B. D. 2017, Phys. Rev. Lett., 119, 1102

\bibitem[Siegel \& Metzger(2018)]{siegel18} Siegel, D. M., Metzger, B. D. 2018, ApJ, 858, 52

\bibitem[Tanaka \& Hotokezaka(2013)]{tanaka13} Tanaka, M. \& Hotokezaka, K. 2013, ApJ. 775, 113  

\bibitem[Tanaka et al.(2014)]{tanaka14} Tanaka, M., Hotokezaka, K., Kyutoku, K., Wanajo, S., Kiuchi, K., Sekiguchi, Y., Shibata, M. 2014, ApJ, 780, 31

\bibitem[Wang \& Wheeler(2008)]{wang08} Wang, L. F. \& Wheeler, J. C., 2008, Annu. Rev. Astron. Astrophys., 46, 433

\bibitem[Wanajo et al. (2014)]{wanajo14} Wanajo, S., Sekiguchi, Y., Nishimura, N., Kiuchi, K., Kyutoku, K., Shibata, M. 2014, ApJ, 789, L39

\end{thebibliography}
\end{document}